# Quantitative study of Silicon Waveguides for the Generation of Quantum Correlated Photon Pairs Bridging Mid-Infrared and Telecom Bands

**Abhishek Kumar Pandey[1], Deepak Jain[1] and Catherine Baskiotis[2],***

[1] Optics and Photonics Centre, Indian Institute of Technology Delhi, New Delhi 110016, India
[2] De Vinci Research Centre, 12 avenue Léonard de Vinci, 92916 Paris La Défense, France

*Author to whom any correspondence should be addressed.

**E-mail:** opz228651@opc.iitd.ac.in and catherine.baskiotis@devinci.fr

**Abstract**

Sources of quantum correlated photons pairs bridging the 3μm-4μm Mid-infrared (MIR) band and Telecom/Near-Infrared/Visible band are of high importance for quantum technologies. Spontaneous Parametric Down Conversion is generally used for realizing such sources, but requires costly implementation platforms with reduced versatility. Here, we explore the potentialities of Spontaneous Four-Wave Mixing (SFWM) in all-solid Silicon On Insulator (SOI) waveguides thanks to an experimentally validated model and propose designs ensuring the production of correlated photon pairs bridging the 3μm-4μm Mid-infrared band and Telecom C-band. Choosing a pump with a wavelength in the range 2100nm-2210nm and a pulse duration of 5ps, we quantitatively performed simulations targeting a probability of photon pair generation per pulse of 0.05, and we found realistic conditions of utilization (2cm-length straight waveguides, intra-modal Four Wave Mixing with the fundamental $TE_{00}$ mode) with a pump peak power in between 9.2mW and 32mW. A first design (wCOM) reaches a signal wavelength as high as 3.905μm, which is situated in an atmospheric transparency window, while maintaining an idler in the Telecom C-band, making it of high interest for atmospheric Quantum Key Distribution. Two other designs wCH$_4$ and wNO$_2$ aim precise CH$_4$ and NO$_2$ gas sensing with a signal wavelength of 3265nm and 3461nm respectively. In terms of signal/idler wavelength separation, wCOM attains the value of 2364 nm which is well above the current record of ~1125 nm obtained in quantum regime with SFWM in all-solid SOI waveguides.

## 1. Introduction

Sources of quantum correlated single-photon pairs with large wavelength spacing bridging 3μm-4μm Mid-infrared (MIR) band and Telecom/Near-Infrared/Visible band are especially attractive for numerous applications [1]. First of all, these sources present high potential for transferring current single-photon daylight atmospheric Quantum Key Distribution (QKD) schemes from Telecom band [2] to the MIR, which is an identified improvement strategy for overcoming present limitations [3]. Indeed, smaller constituent gas absorption and reduced Rayleigh scattering [4] are observed in the in the 3μm-4μm MIR band as compared to the Telecom band (cf. figure 1.a), as well as lower solar noise and increased resistance to atmospheric meteorological variations [5,6]. Secondly, these sources can serve in second order interferences experiments [7-9], to realize "sensing with undetected light" schemes or "imaging with undetected light" schemes in which sensing or imaging in the MIR is realized using only visible/Near-infrared/telecom detectors [8,9]. Such schemes have been theoretically and experimentally introduced, in 1991, in the frame of photon-level sensing [7]. Then, one year later, in 1992, in the reference [10], authors theoretically predicted that sensing with undetected light can be achieved with classical light and not only photon-level sensing. The first experimental demonstration of Belinsky *et al*. theoretical predictions aroused in 2016 with the work of Kalashnikov *et al*. [8]. Since that publication, several works explored the potential of such schemes both in the case of classical light [9-13] and in the case of photon-level sensing/imaging [14]. It is worth noting that photon-level MIR gas sensing enables measurements of very low concentration with high precision [15], thanks to the combination of a single-photon sensing and a sensing in the MIR, which is the fingerprint region of numerous gases such as NO$_2$ and CH$_4$ [16], cf. figure 1.(b). In all literature references, high-$\chi^2$ crystals are used in bulk-optics open-air experiments for producing the photon pairs. Replacing these bulk-optics high-$\chi^2$ crystals by ultra-compact integrated waveguides is of high interest in terms of versatility, cost, compactness and lightness. Lightness and compactness are of special interest for atmospheric QKD, where weight and dimension reduction of the satellite payload is critical. On their side, cost and versatility are of high concerns for all the above-mentioned applications. Additional last noteworthy fact, most of the reference are reporting correlated pairs connecting MIR band and Visible/[Near Infrared] band, while a strong interest in connecting MIR band and Telecom band can be anticipated thanks to the commercial availability, in the Telecom band, of single photon detectors together with the existence of mature light processing devices and equipment, that are often not adapted or completely unexploitable in the visible. An example of utilization of Telecom band equipment can be found in reference [17], where a temperature-controlled 10km-Single Mode Fibre (SMF) is used to realize dispersive Fourier Transformation to perform mid-IR spectroscopy. Moreover, with the telecom band, eventually, all optical information processing becomes possible, which will be of dramatic interest for quantum QKD.

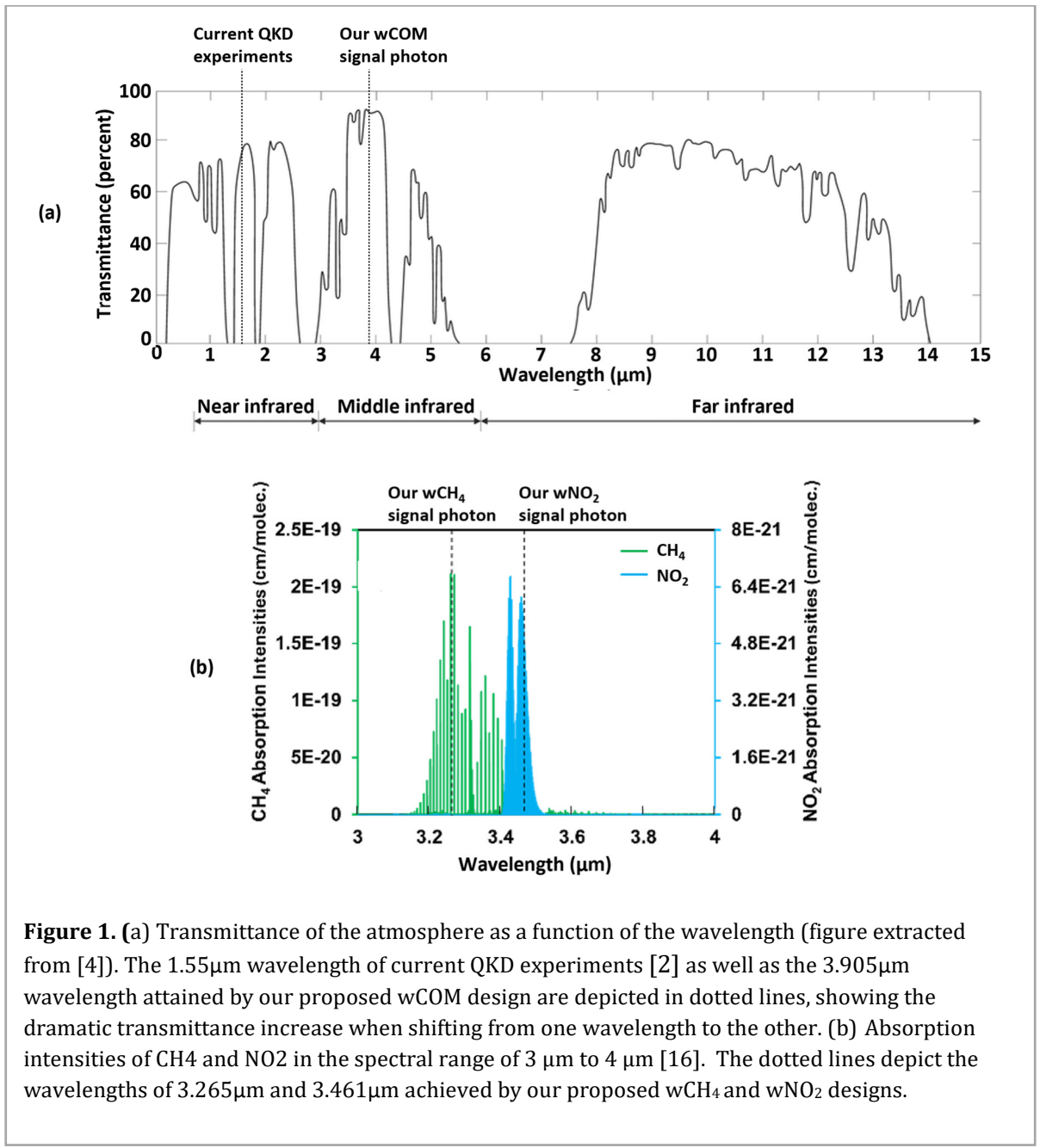

**Figure 1.** (a) Transmittance of the atmosphere as a function of the wavelength (figure extracted from [4]). The 1.55μm wavelength of current QKD experiments [2] as well as the 3.905μm wavelength attained by our proposed wCOM design are depicted in dotted lines, showing the dramatic transmittance increase when shifting from one wavelength to the other. (b) Absorption intensities of CH4 and NO2 in the spectral range of 3 μm to 4 μm [16]. The dotted lines depict the wavelengths of 3.265μm and 3.461μm achieved by our proposed wCH$_4$ and wNO$_2$ designs.

In this work, we exploit the large $\chi^3$ parameter of Silicon waveguides to produce, by Spontaneous Four Wave Mixing (SFWM), correlated photon single-pair bridging 3μm-4μm MIR band and Telecom band. Silicon waveguides are highly scalable and suitable for practical use in integrated quantum devices thanks to their easy and cheap fabrication, their compactness, as well as their compatibility with existing CMOS technologies [18,19]. Previous studies have experimentally reported the possibility of producing, by SFWM in air-clad Silicon waveguides, correlated beams with large gap between signal and idler (~2058nm), in the case of a classical regime in which the pump peak power is tuned to a large value (20W) in order to achieve phase matching [20]. Despite significant losses induced by surface roughness on the three air-clad core facets [21], these designs appear therefore especially promising. However, in the frame of a quantum regime, in which only one photon-pair has to be emitted at maximum at each pump pulse, it is required to keep a low value for the pump peak power, which, therefore, cannot be exploited to realize the phase-matching. With the same designs, the achievement of phase matching at near zero pump peak power is performed at the cost of a dramatic signal/idler wavelength gap reduction to 1780nm, with an idler wavelength of 1610nm in the L-band and a signal wavelength of 3423nm outside any atmospheric transmission window, cf. figure 1.(a). It follows that, previous designs proposed in [20] cannot be used for atmospheric QKD in single-pair regime. In addition, beyond phase matching achievement, the existing literature lacks a quantitative study assessing the actual feasibility of producing correlated photon pairs in the quantum regime. As a result, it remains uncertain whether such sources can indeed generate correlated photon pairs in the quantum regime.

In this paper, we theoretically demonstrate that it is possible to use all-solid silicon integrated waveguides to produce, in the quantum regime, correlated photon pairs bridging 3μm-4μm Mid-infrared band and Telecom C-band. On the contrary of previous papers [20,22], which are not precisely studying the quantum regime, we provide a quantitative prediction of the probability of photon pair generation per pulse founded on a model that we experimentally validated. We propose three designs targeting gas sensing and atmospheric QKD. The design targeting atmospheric QKD reaches signal-idler wavelength separation up to ~2364 nm, which is substantially larger than the record value (~1125 nm) theoretically reported with SFWM in earlier all-solid Silicon-waveguide studies [22]. The signal lies near 3905nm within MIR atmospheric transparency window, while the idler remains in the telecom C-band. This paper is organized in four sections. The section 2 provides detailed information about our modeling methods of SFWM generation of correlated photon pairs, by indicating the precise formula that we used for the calculation of the probability of photon pair generation and depicting its experimental validation. The section 3 presents our three proposed designs for correlated photon pairs that bridge the MIR band and telecommunications band. Finally, the section 4 discusses the several design choices as well as the potential applications for sensing and atmospheric QKD.

## 2. Modeling the generation of correlated photon pairs through SFWM in Silicon waveguides

In this section, we first present the study choices, then the guided modes modeling of our all-solid Silicon waveguides, then the nonlinear parameter calculation, then the energy and momentum conservation conditions, then the modeling method for the calculation of the probability of photon pair generation, and finally the model experimental validation.

### 2.1. Study choices

We studied the generation of correlated photon pairs by degenerated SFWM in an intramodal arrangement within a silicon waveguide. The two pump photons exhibit the same frequency $\omega_P$ and are annihilated to create a pair of photons with one idler photon at frequency $\omega_i$ and one in the $TE_{00}$ mode, cf. Figure 2.(a). The chosen geometry is an all-solid step-index Silicon-On-Insulator waveguide which consists of a Silicon (Si) core deposited on a buried oxide (BOX) layer, surrounded by a fused silica (SiO2) cladding, cf. Figure 2.(b). We chose to study the case of a pulsed pump with a hyperbolic secant temporal profile.

### 2.2. Guided modes modeling

For determining the characteristics of the effective indices and the spatial profiles of the fundamental $TE_{00}$-guided modes of Silicon waveguides, we utilized the COMSOL Multiphysics's Wave Optics Module which implements the Finite Element Method (FEM). The refractive indices of the core and cladding were determined using Sellmeier's equations [23-26] while the BOX layer refractive index is chosen identical to that of SiO2. Simulations allowed us to compute the effective indices, the propagation constant and the mode profile of the guided modes.

### 2.3. Nonlinear parameter determination

The nonlinear coefficient of the silicon waveguide associated with SFWM is given by [27]:

$$\gamma = \frac{2\pi n_2 f_{ppsi}}{\lambda_p} \quad (1)$$

$n_2$ represents the nonlinear refractive index of silicon (which is obtained from [28]), $\lambda_p$ represent the wavelength of the pump, $f_{ppsi}$ is the overlap integral for the $TE_{00}$ mode at the pump, signal, and idler wavelengths [27]:

$$f_{ppsi} = \frac{\iint E_p^* E_p^* E_s \, E_i \, dxdy}{\left(\iint |E_p|^2 dxdy \iint |E_p|^2 dxdy \iint |E_s|^2 dxdy \iint |E_i|^2 dxdy\right)^{1/2}} \quad (2)$$

where the double integrals are performed over the entire waveguide cross-section and $E_p, E_s, E_i$ are the transverse mode field of the $TE_{00}$ mode at the pump, signal, and idler wavelengths, respectively.

### 2.4. Energy and momentum conservation conditions

A quantum mechanical study on SFWM in the waveguide shows that the probability of generating a photon pair is substantial only in the vicinity of the fulfillment of both the momentum conservation condition (also called the Phase-Matching Condition) and the energy conservation condition [29,30]:

$$2\beta(\omega_P) - \beta(\omega_s) - \beta(\omega_i) = 0 \quad (3)$$

$$2\omega_P = \omega_s + \omega_i \quad (4)$$

correlated photon pair per pulse, the peak pump power is very low, and we therefore neglected in equation (3) the additional phase coming from the Self Phase Modulation (SPM) of the pump.

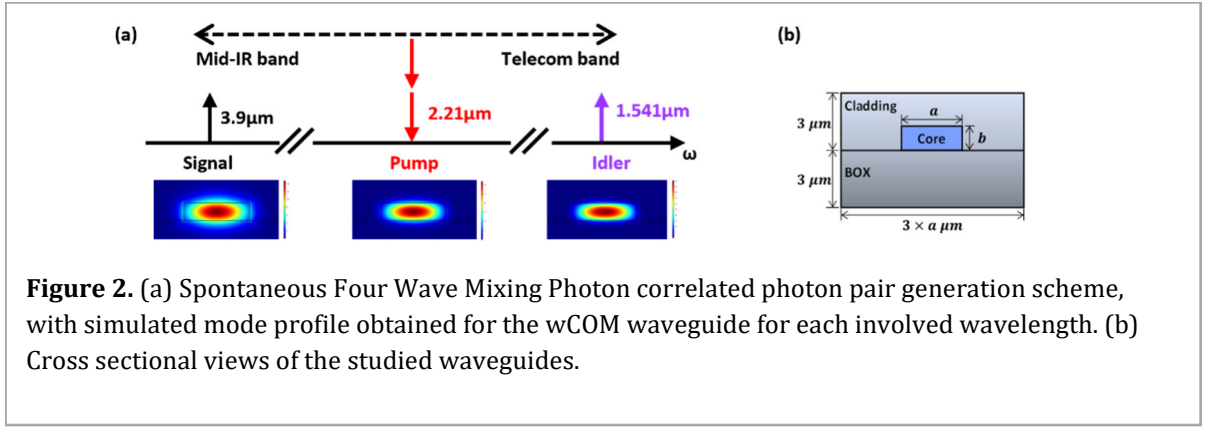

**Figure 2.** (a) Spontaneous Four Wave Mixing Photon correlated photon pair generation scheme, with simulated mode profile obtained for the wCOM waveguide for each involved wavelength. (b) Cross sectional views of the studied waveguides.

### 2.5. Photon Pair Generation per Pulse (PGP)

For the modeling of the probability of photon Pair Generation per Pulse ($PGP$), we used the model outlined in Margaux Barbier's thesis [30]. This model has been derived under the assumptions of negligible losses in the Silicon waveguide and a narrow spectral extension of the pump. We first present the model for the calculation of the Joint Spectral Density, then the model for the calculation of the probability of the photon Pair Generation per Pulse, then an experimental validation of the $PGP$ calculation using experimental results that have been previously reported in the literature [31].

### 2.5.1 Joint Spectral Density (JSD)

The probability of generating a pair of correlated photons, with one photon in the spectrum range $[\omega_s - d\omega_s/2;\ \omega_s + d\omega_s/2]$ and the other in the spectrum range $[\omega_i - d\omega_i/2;\ \omega_i + d\omega_i/2]$ during the propagation of the pump pulse in the waveguide, is expressed as [31]:

$$Prob(\omega_s, \omega_i) = |\zeta_{2D}(\omega_s, \omega_i)|^2 d\omega_s d\omega_i \qquad (5)$$

where is $|\zeta_{2D}(\omega_s, \omega_i)|^2$ the two-dimensional joint spectral density (JSD) of the correlated photon pair generation probability. Under the assumptions of narrow spectral extension of the pump, and negligible losses in Silicon waveguide, $|\zeta_{2D}(\omega_s, \omega_i)|^2$ can be expressed as a product of three factors [31]:

$$|\zeta_{2D}(\omega_s, \omega_i)|^2 = A(\omega_s, \omega_i) \times G(\omega_s, \omega_i) \times F(\omega_s, \omega_i) \qquad (6)$$

The factor $A(\omega_s, \omega_i)$ is given by:

$$A(\omega_s, \omega_i) = \left(\frac{n_{gp}\sqrt{n_{gs}n_{gi}}}{2\pi}\sqrt{\frac{\omega_s\omega_i}{\omega_p^2}}\gamma L \mathcal{E}_{imp}\right)^2 \qquad (7)$$

where:
- $n_{gp}$, $n_{gs}$ and $n_{gi}$ are the group indices of the pump, signal and idler photons;
- $L$ is the length of the waveguide;
- $\mathcal{E}_{imp}$ is the energy of a pump pulse.

In the case of a hyperbolic secant temporal profile for the pump pulse, $\mathcal{E}_{imp} = 2P_P T_0$, with $P_P$ the peak power and $T_0$ the duration of the pump pulse.

The factor $G(\omega_s, \omega_i)$ is given by (still considering hyperbolic secant temporal profile for the pump pulse) [31]:

$$G(\omega_s, \omega_i) = \left(\frac{\frac{\pi T_0}{2}x}{\sinh\left(\frac{\pi T_0}{2}x\right)}\right)^2 \qquad (8)$$

where $x$ is the phase mismatch and is given by $x = \omega_s + \omega_i - 2\omega_P$.

The factor $F(\omega_s, \omega_i)$ accounts for phase-matching conditions and dispersion effects. It is expressed as:

$$F(\omega_s, \omega_i) = sinc^2\left[\left(2\beta(\omega_P) - \beta(\omega_s) - \beta(\omega_i) + \frac{x}{v_{gp}}\right)\frac{L}{2}\right] \quad (9)$$

where $v_{gp}$ is the group velocity of pump photons.

### 2.5.2 Probability of photon Pair Generation per Pulse (*PGP*)

Reminding that, in experiments, filters are placed in front of the signal and idler single-photon detectors, and denoting $\Delta\omega_i$ and $\Delta\omega_s$ the bandwiths of the idler and signal filter respectively, we define the Probability of photon pair Generation per Pulse (*PGP*) as the probability of generation of a pair with an idler photon lying in the frequency range $\left[\omega_i - \frac{\Delta\omega_i}{2}; \omega_i + \frac{\Delta\omega_i}{2}\right]$ and a signal photon lying in the frequency range $\left[\omega_s - \frac{\Delta\omega_s}{2}; \omega_s + \frac{\Delta\omega_s}{2}\right]$.

We first computed the one-dimensional Spectral Density of Probability of photon pair generation $|\zeta_{1D}(\omega_s)|^2$ at the frequency $\omega_s$, performing the integral of the bidimensional spectral density $|\zeta_{2D}(\omega_s, \omega_i)|^2$ over all possible collected partner $\omega_i$ frequencies:

$$|\zeta_{1D}(\omega_s, \omega_i)|^2 = \int_{\omega_i - \frac{\Delta\omega_i}{2}}^{\omega_i + \frac{\Delta\omega_i}{2}} |\zeta_{2D}(\omega_s, \omega_i)|^2 \, d\omega_i \quad (10)$$

We then obtained $PGP(\omega_s, \omega_i)$ by performing the integration of $|\zeta_{1D}(\omega_s, \omega_i)|^2$ over all possible collected $\omega_s$ frequencies [31]:

$$PGP(\omega_i) = \int_{\omega_s - \frac{\Delta\omega_s}{2}}^{\omega_s + \frac{\Delta\omega_s}{2}} |\zeta_{1D}(\omega_s, \omega_i)|^2 \, d\omega_s \quad (11)$$

### 2.5.3 Probability of photon Pair Generation per Pulse (*PGP*)

To experimentally validate the formula (11), we used the experimental results presented in the publication of Alibart *et al.* [31]. Although this publication studies SFWM in Photonic Crystal Fibres (PCF) and not integrated waveguides, it is the only one that we found in the literature which reports experimental correlated photon pair generation rate, in pulsed regime, together with a precise description of the characteristics of the waveguide and the set-up used for the experiments. In this publication, 0.2m of PCF are used to generate correlated photon pairs with a photon in the visible and a photon in the near infrared (cf. Figure 3).

Following the path of the authors, we modelized the PCF by a simple step-index fibre. We consider a fused silica core with a refractive index computed through standard Sellmeier coefficients [25]. To determine the refractive index of the cladding we compute the mean refractive index obtained considering a 90% air-filling fraction. We found out the characteristics of the fundamental guided mode of the considered step-index fibre with large refractive index contrast employing a standard full vectorial model [32].

We considered a core radius of 0.965µm, which is coherent with the indication of the publication [31] according to which the core diameter of the PCF is of ~2µm. All the other simulation parameters are directly taken from the publication [31]. We therefore set the pump wavelength to 708.4nm, the pulse duration to $\tau_{PCF} = 2ps$, the repetition rate to $R = 80MHz$, the central wavelengths of the collecting filters to 570nm and 880nm, and the bandwidth of the filter to 40nm. In terms of frequencies, we therefore obtained, for the filter, central frequencies of 526 THz and 341 THz with a bandwidth of 36.9 THz and 15.5 THz respectively. For computing the overlap integral $f_{ppsi}$, we performed the approximation: $f_{ppsi} = 1/A_{eff}$, where $A_{eff}$ is the effective area of the pump mode obtained using a full vectorial formula.

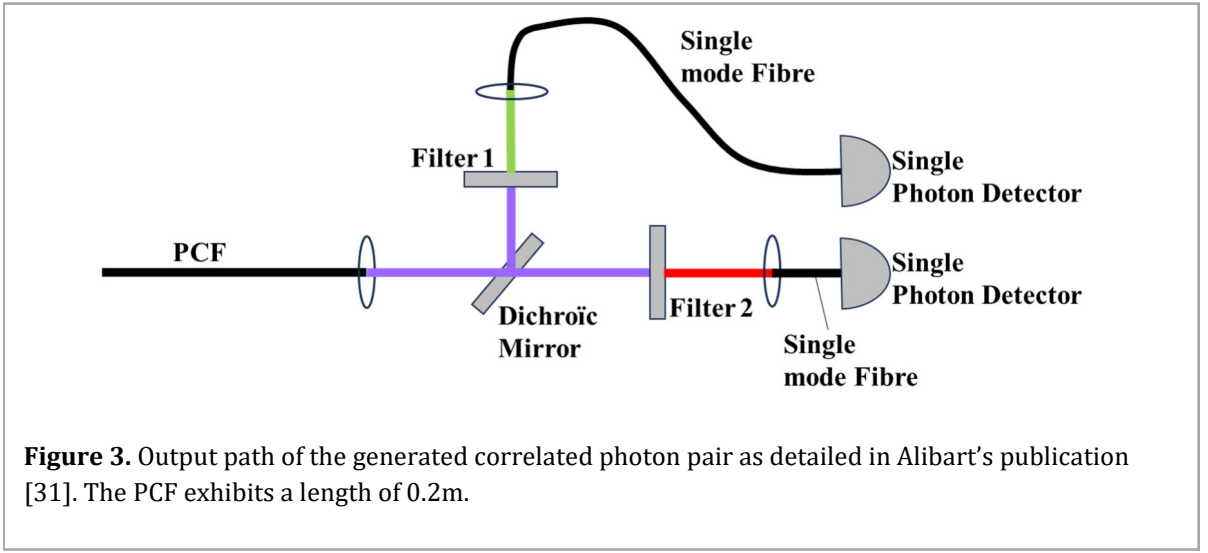

**Figure 3.** Output path of the generated correlated photon pair as detailed in Alibart's publication [31]. The PCF exhibits a length of 0.2m.

We computed the Photon Pair Generate Rate ($PGR$) by simply multiplying the $PGP$ by the repetition rate of the pump $R$:

$$PGR = PGP \times R. \tag{12}$$

Considering the optical path of the photons after their generation (cf. Figure 3), we took account of the attenuation of the different components of the experimental set-up by multiplicating the obtained $PGR$ with an attenuation factor $A_{attenuation}$ and we obtained the theoretical photon pair rate:

$$r_{Barbier} = A_{attenuation} \times PGR \tag{13}$$

For computing $A_{attenuation}$, we took an overall coupling efficiency from the PCF to the collecting single mode fibre of $\mu_s \approx 0.58$ and $\mu_i \approx 0.44$ for the signal and the idler respectively, and a Single Photon Detector efficiency of $\eta_s \approx 0.60$ and $\eta_i \approx 0.33$ for the signal and idler respectively, following the indications of the authors of the publication [31].

Considering the optical path of the photons after their generation, we determined:

$$A_{attenuation} = 0.58 \times 0.44 \times 0.60 \times 0.33. \tag{14}$$

We deduced the pump peak power $P_P$ from the mean pump power $P_{mean}$ indicated in the reference [31] considering a gaussian pulse approximation:

$$P_P = \frac{2}{R \times \tau_{PCF}} \sqrt{\frac{\ln(2)}{\pi}} P_{mean}. \tag{15}$$

Table 1 indicates the results that we obtained by computing the theoretical photon pair rate $r_{Barbier}$ for different launched mean pump power into the PCF fibre, as well as the experimentally measured photon pair rate $r_{exp}$ presented in Alibart's publication [31]. We also indicated the theoretical predictions of the photon pair rate $r_{th}$ obtained in Alibart's publication. We can observe a very good agreement between the results obtained with the model used in this paper ($r_{Barbier}$) and the experimental results ($r_{exp}$) and a better accuracy of the Barbier's model as regards to the model proposed in Alibart's publication. We analyze the gain in accuracy achieved with the Barbier's model, by the fact that Alibart's model has been derived though the application of numerous approximations. The main difference between the two models resides in the fact that, in Barbier model, the spectral extent of the pump is taken into account through the factor $G(\omega_s, \omega_i)$ of the equation (8) and the additional term $(x/v_{gp}) \times (L/2)$ in the argument of the $x \rightarrow sinc(x)$ function, while in Alibart's model the spectral extent of the pump is taken into account through an approximate formula valid for gaussian pulses.

**Table 1.** Validation of the formula (13) and a fortiori (11) using the results reported in Alibart's publication [31]. From left to right, the mean pump power launched into the PCF, the photon pair rate $r_{Barbier}$, the experimentally measured photon pair rate $r_{exp}$ obtained in Alibart's publication and the theoretical photon pair rate $r_{th}$ obtained in Alibart's publication.

| Mean pump Power (µW) | $r_{Barbier}$ ($s^{-1}$) | $r_{exp}$ ($s^{-1}$) | $r_{th}$ ($s^{-1}$) |
|---|---|---|---|
| 960 | $8.70 \times 10^6$ | $8.46 \times 10^6$ | $8.05 \times 10^6$ |
| 660 | $4.29 \times 10^6$ | $4.08 \times 10^6$ | $3.81 \times 10^6$ |
| 490 | $2.42 \times 10^6$ | $2.31 \times 10^6$ | $2.10 \times 10^6$ |
| 340 | $1.19 \times 10^6$ | $1.14 \times 10^6$ | $1.10 \times 10^6$ |
| 200 | $4.20 \times 10^5$ | $0.43 \times 10^6$ | $0.35 \times 10^6$ |

## 3. Proposed sources of correlated photon pairs bridging MIR and Telecom band
### 3.1 Proposed designs characteristics

We selected three distinct silicon waveguide designs, with the corresponding parameters detailed in Table 2. For each waveguide, we set the waveguide length to 2 cm to optimize phase matching and minimize propagation losses. The Figure 4.(a) shows the evolution of the effective indices of the $TE_{00}$ mode of the three selected waveguides as a function of the wavelength, together with the material dispersion curve. The effective index curves are explicitly accounting the combined effect of material and waveguide dispersion through the inclusion of material dispersion in the full-vectorial eigenmode simulations, (cf. section 2.2). Figure 4.(a) reveals that the material dispersion is not dominating. Although, the waveguides possess relatively large core dimensions, the strong confinement introduced by the small height of the waveguide allows precise tailoring of the effective index curve over a wide spectral range. This enables the fulfillment of the phase-matching condition at the targeted pump, signal, and idler wavelengths, even for spectrally far-separated photon pairs. The Figure 4.(b) shows the phase-matched idler and signal wavelength as a function of the pump wavelength for the three designs. On this figure, we can read for each pump wavelength the signal and idler wavelengths corresponding to fulfillment of the equation (3). Waveguides wCH$_4$ and wNO$_2$ are intended to gas sensing. We designed the wCH$_4$ waveguide such as the signal photon is emitted within an absorption band of CH$_4$ [16], while being out of any NO$_2$ absorption bands [16], cf. Figure 1.(b). In the same way, we optimized the wNO$_2$ waveguide such as the signal photon belongs to a NO$_2$ absorption band, while being out of any CH$_4$ absorption bands. Finally, we designed the waveguide wCOM such that the simulations predict signal photons emission at 3.905µm, namely within an atmospheric transparency window in such a way to facilitate efficient free-space QKD [3]. In our three optimized designs, the idler photon is generated in the telecommunications C-band (1530–1565 nm). Red diamonds on the Figure 4.(b) mark the chosen pump, idler and signal wavelengths. We found out the three designs by performing a grid search to explore the possibility of phase matching condition fulfillment for the targeted wavelengths.

**Table 2.** Selected waveguides characteristics.

| Parameter | wCH$_4$ | wNO$_2$ | wCOM |
|---|---|---|---|
| Core Dimension $a(\mu m) \times 0.75(\mu m)$ | 2.05 × 0.75 | 2.23 × 0.75 | 2.35 × 0.65 |
| MIR Application | CH$_4$ Detection | NO$_2$ Detection | Atmospheric QKD |
| Peak Pump Power (mW) | 24.1 | 9.2 | 32.2 |
| Pump Wavelength μm | 2.100 | 2.151 | 2.210 |
| Signal Wavelength μm | 3.265 | 3.461 | 3.905 |
| Idler Wavelength μm | 1.547 | 1.560 | 1.541 |
| Overlap Integral (m$^{-2}$) | $1.05 \times 10^{12}$ | $2.90 \times 10^{12}$ | $9.69 \times 10^{11}$ |
| Nonlinear Parameter (W$^{-1}$m$^{-1}$) | 20.78 | 55.74 | 16.66 |

For all the three designs, the simulated idler and signal wavelengths belong to the Silicon transparency window with material losses of approximately 0.1 dB/cm and 0.002 dB/cm, respectively [33]. The phase matching curve for each waveguide as a function of the pump and signal/idler wavelengths are depicted in Figure 4.(b). Using the overlap integral calculation (Eq. 2) and the literature values of the nonlinear refractive index of silicon [28] we calculated the nonlinear coefficient ($\gamma$) for each waveguide (Eq. 1), cf. Table 2. After choosing a pump pulse duration of 5ps, we plot the Normalized Joint Spectral Density (JSD), cf. Figure 5.(a-c). We considered 1 THz bandwith for the signal and idler filters positioned just before the detectors. The probability of photon Pair Generation per Pulse $PGP$ (as computed with the Eq. 11) is depicted in Figure 4.(c). For the three waveguides, the idler wavelength falls in the Telecom band and can be detected in a Dense Wavelength Division Multiplexing (DWDM) International Telecommunication Union (ITU) channel, making the detection easy with telecom equipment.

The wavelengths at which occurs the peak $PGP$ for the signal and the idler photon are reported in Table 2. In particular, we attained a signal wavelength of 3.265μm, 3.461μm and 3.905μm for the waveguides wCH$_4$, wNO$_2$, and wCOM respectively. We choose to target a peak $PGP$ value of ~0.05 by considering previous experimental studies, which have shown that operating at a photon-pair generation probability per pulse in the range of approximately 0.01 - 0.1 provides a practical compromise between source brightness and suppression of multi-pair generation events [1, 30, 34]. We selected the target value of PGP ≈ 0.05 as a representative operating point within the well-reported range. This choice is not a fundamental limitation of the proposed designs. Since, according to Eq. 7, the $PGP$ scales with the square of the pump peak power $P_P$ and the waveguide length $L$: $(P_P \times L)^2$, higher $PGP$ values could be achieved if required by increasing the pump power $P_P$.

### 3.2 Tolerance to small fabrication imperfections

We numerically investigated the tolerance of our three proposed waveguides to fabrication imperfections by slightly varying the core dimensions. Specifically, we studied four cases. We changed the width a of ± 0.01 μm, keeping the height b fixed. We also changed the height b of

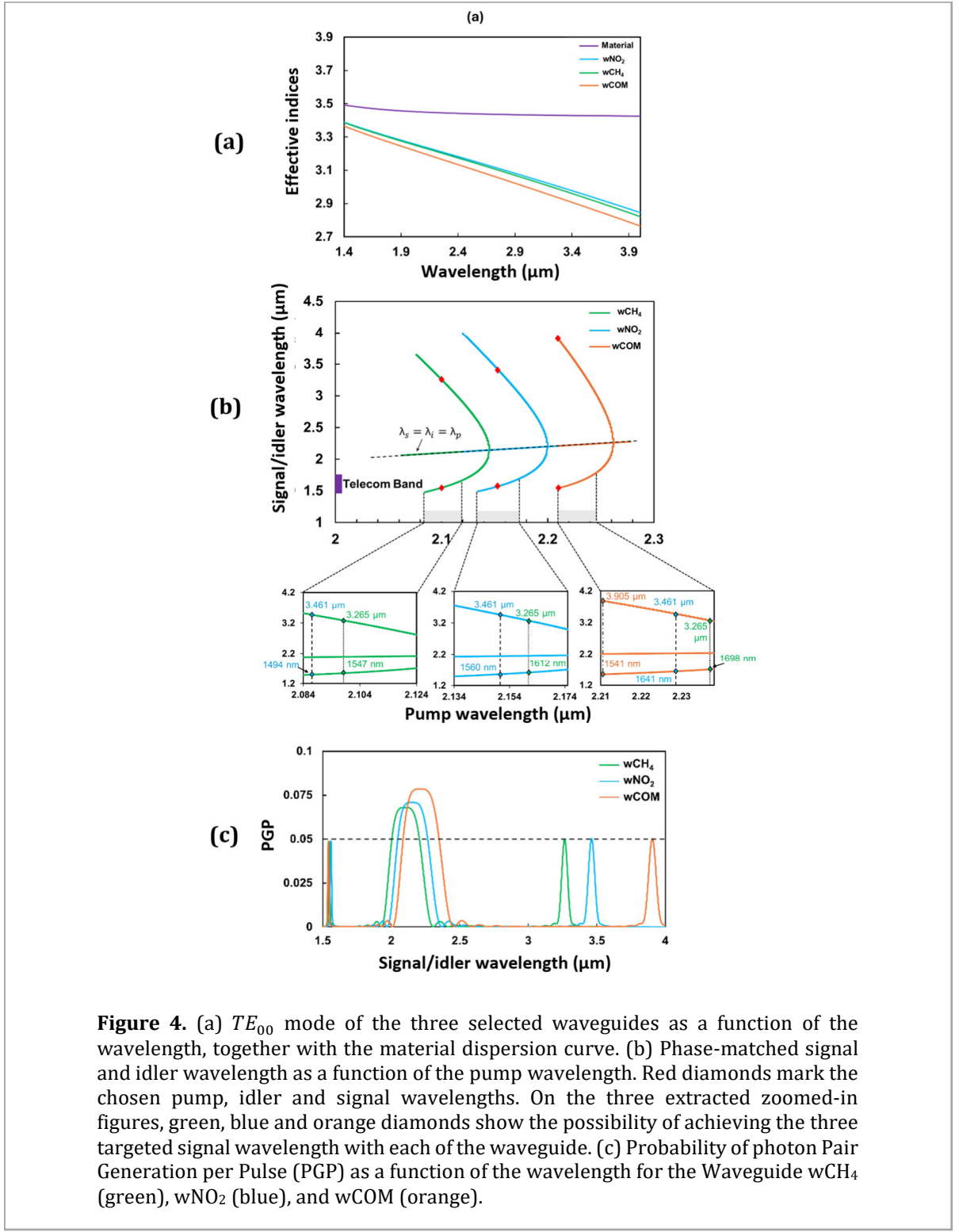

**Figure 4.** (a) $TE_{00}$ mode of the three selected waveguides as a function of the wavelength, together with the material dispersion curve. (b) Phase-matched signal and idler wavelength as a function of the pump wavelength. Red diamonds mark the chosen pump, idler and signal wavelengths. On the three extracted zoomed-in figures, green, blue and orange diamonds show the possibility of achieving the three targeted signal wavelength with each of the waveguide. (c) Probability of photon Pair Generation per Pulse (PGP) as a function of the wavelength for the Waveguide wCH$_4$ (green), wNO$_2$ (blue), and wCOM (orange).

± 0.01 µm, keeping the width a constant (cf. Figure 6). For waveguides wCH$_4$ and wNO$_2$, for small variation of the height (b ± 0.01 µm) with fixed width a, no drift is observed in the signal wavelength, cf. Figure 6.(c,g,d,h). For all other cases, we observed a small drift of the signal and idler wavelength. Nonetheless, we compensated this drift by slightly changing the pump wavelength, cf. Figure 6.(c,g,d,h). For all other cases, we observed a small drift of the signal and idler wavelength. Nonetheless, we compensated this drift by slightly changing the pump wavelength of a few nanometers in such a way to obtain peak **PGP** at the initially targeted signal wavelengths indicated in Table 2. Performing this change, we observed that the idler photon remains in the C-band cf. Figure 6.(a,b,e,f,i-l). In practice, the pump wavelength can be tuned using a tunable laser source such as an Optical Parametric Oscillator (OPO), which allows access to a broad range of wavelengths, including 2.0–2.236µm with a simple temperature tuning [35]. Our study aligns of a few nanometers in such a way to obtain peak ***PGP*** at the initially targeted signal wavelengths indicated in Table 2. Performing this change, we observed that the idler photon remains in the C-band cf. Figure 6.(a,b,e,f,i-l). In practice, the pump wavelength can be tuned using a tunable laser source such as an Optical Parametric Oscillator (OPO), which allows access to a broad range of wavelengths, including 2.0–2.236µm with a simple temperature tuning [35]. Our study aligns with prior works in Silicon photonics, which study ±10 nm variation in waveguide dimensions [36]. Our results therefore predict a good robustness of our proposed waveguides to small fabrication imperfections.

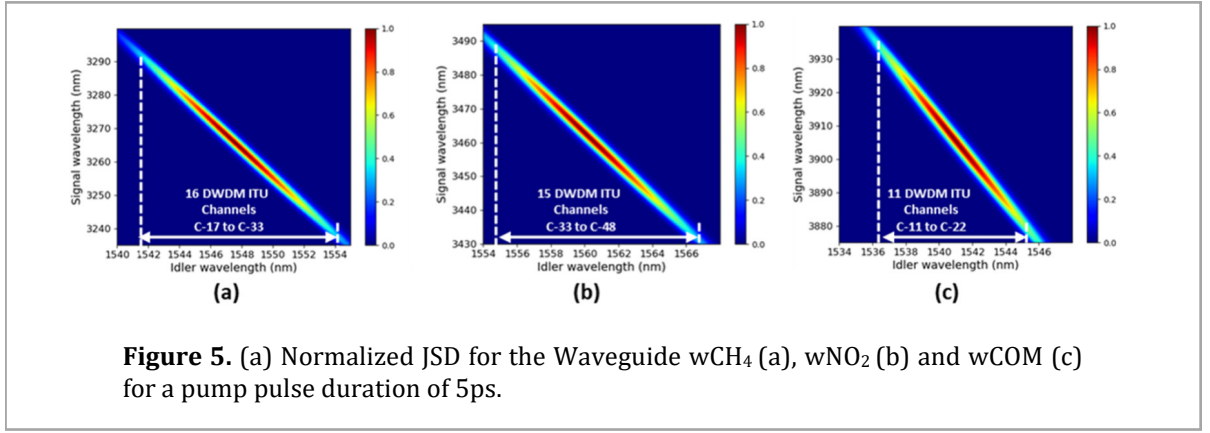

**Figure 5.** (a) Normalized JSD for the Waveguide wCH$_4$ (a), wNO$_2$ (b) and wCOM (c) for a pump pulse duration of 5ps.

### 3.3 Potentialities of the waveguides to be used for a large range of signal wavelength

From the Figure 3.(b) showing the phase matching curves, we extracted three zoomed-in figures corresponding to an idler wavelength falling in the enlarged Telcom band. On these extracted figures, green, blue and orange diamonds show the possibility of achieving the three targeted signal wavelength with each of the waveguide. These figures reveal that if we impose the condition that the idler has to remain in the Telecom C-band, none of the waveguide can reach the wavelength targeted by the others. Nonetheless, if we extend the idler range to the whole enlarged Telecom band, the waveguide wCOM can be used for the three targeted wavelengths. Depending on the experimental setup, either Telecom C-band devices are required, making it necessary to use three different waveguides, either devices that operate over an extended Telecom band can be used, allowing the use of only one waveguide.

## 4. Discussions and applications

### 4.1 Implications of the choice of intramodal SFWM

We chose to use intramodal SFWM, which presents numerous assets. In intramodal SFWM, all interacting waves go within the same mode of the waveguide, hence reducing the effects of modal dispersion and temporal walk-off between modes observed in intermodal systems [37]. The mode uniformity increases the spectrum overlap between pump, signal and idler mode, leading to improved photon-pair generation efficiency as compared to intermodal SFWM [29]. It is worth noting that the previous literature record of (~1125 nm), in terms of signal/idler separation for the generation of photon pairs in Silicon waveguides through SFWM is achieved by an intermodal configurations and not intramodal, which enhance the interest of our proposed designs.

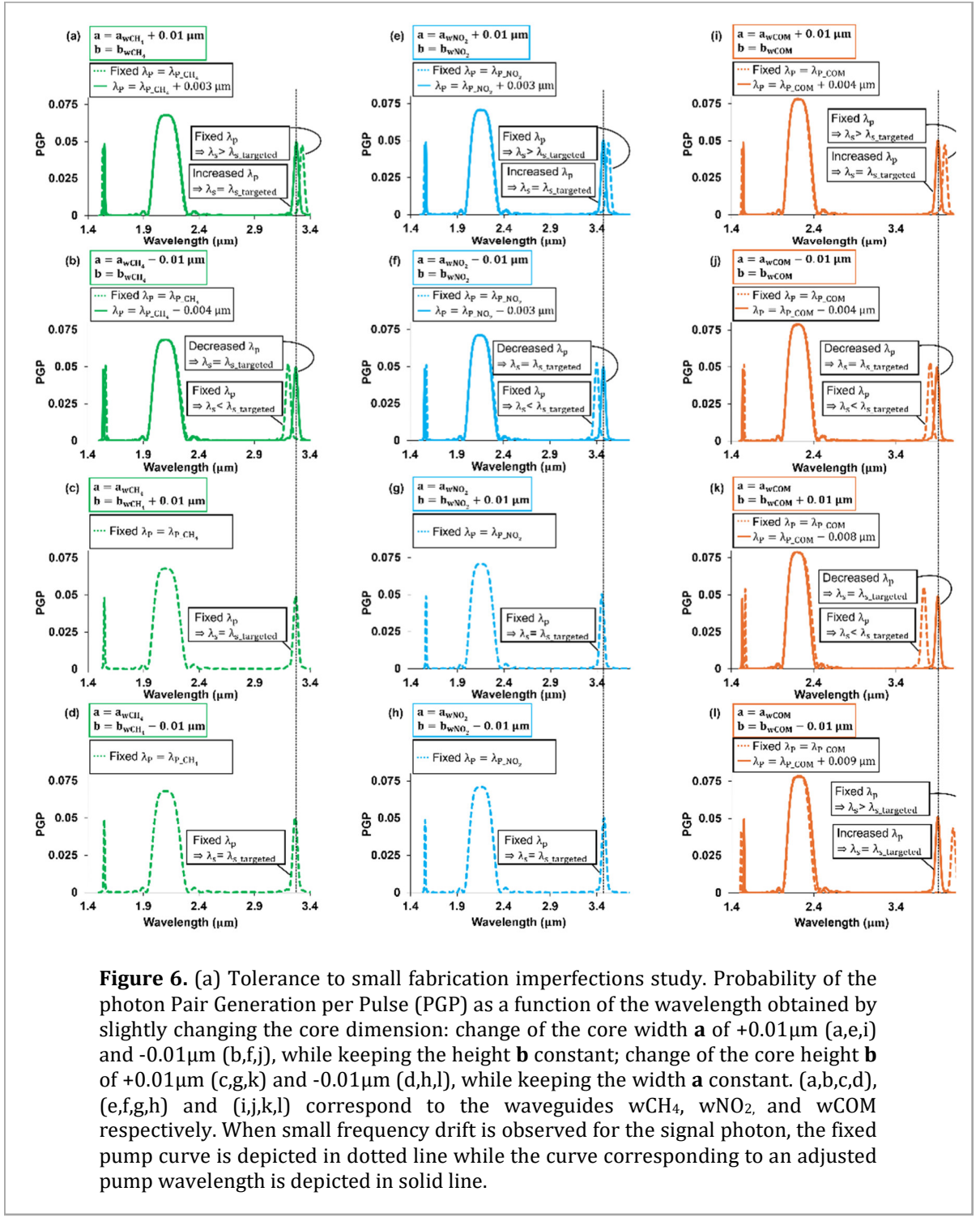

**Figure 6.** (a) Tolerance to small fabrication imperfections study. Probability of the photon Pair Generation per Pulse (PGP) as a function of the wavelength obtained by slightly changing the core dimension: change of the core width **a** of +0.01μm (a,e,i) and -0.01μm (b,f,j), while keeping the height **b** constant; change of the core height **b** of +0.01μm (c,g,k) and -0.01μm (d,h,l), while keeping the width **a** constant. (a,b,c,d), (e,f,g,h) and (i,j,k,l) correspond to the waveguides wCH$_4$, wNO$_2$, and wCOM respectively. When small frequency drift is observed for the signal photon, the fixed pump curve is depicted in dotted line while the curve corresponding to an adjusted pump wavelength is depicted in solid line.

### 4.2 Effect of the material losses and fabrication-induced losses

In our three designs, the signal wavelength is very high in mid-IR, in a region in which the Silica is estimated to be opaque. However, considering the small length of the waveguide and the small percent of power that is propagating in the Silica cladding and in the buried oxide layer, our designs are genuinely functional. The Table 3 depicts the absorption coefficient of fused silica $\alpha_{Silica}$ [38] and Silicon $\alpha_{Silicon}$ [33] at the different signal wavelengths, as well as the percent of power in the core $P_{core,}$, in the fused silica cladding $P_{silica}$, and in the buried oxide layer $P_{BOX}$. Considering that the Buried Oxide layer presents the same material absorption as the fused silica, we computed the total attenuation $A_{material}$ encountered by the signal mode through its propagation into the $L = 2cm$ waveguide by using the formula:

$$A_{material} = P_{core} \times 10^{-\frac{\alpha_{Silicon} \times L}{10}} + (P_{silica} + P_{BOX}) \times 10^{-\frac{\alpha_{Silica} \times L}{10}} \quad (14)$$

**Table 3.** Details of the computation of the total attenuation coefficient $A_{material}$ introduced by material losses at the signal wavelength.

| Wavelength (µm) | $\alpha_{Silica}$ (dB/cm) | $\alpha_{Silicon}$ (dB/cm) | P$_{core}$ (%) | P$_{silica}$ (%) | P$_{BOX}$ (%) | A$_{material}$ |
|---|---|---|---|---|---|---|
| 3.265 | 0.7 | 0.001 | 92.2 | 3.9 | 3.9 | 0.98 |
| 3.461 | 1 | 0.001 | 91.3 | 4.4 | 4.3 | 0.97 |
| 3.905 | 7.3 | 0.002 | 85.3 | 7.4 | 7.3 | 0.86 |

To take account of the propagation losses at the signal wavelength, we multiply PGP by A$_{material}$. Considering that the photon pair generation by SFWM can arise at any point in the waveguide and not only at the beginning, it is important to keep in mind that the new value that we obtain in such way consists in a minimum probability. The idler mode, for his part, is in the Telecom C-band and is therefore encountering negligible losses in Silica cladding. Its impact on the PGP is therefore negligible.

Other losses can be induced by the actual fabrication of the waveguides, as for example, the roughness of the wall of the core or small width difference along the waveguide length. We may remark that the impact of all these losses on the PGP can be experimentally counterbalanced by simply increasing the pump peak power to reach the targeted value.

**4.3 Mitigation of Raman Scattering and Two-Photon Absorption**

Considering that the Raman gain at a specific frequency is only depending on the difference between the considered frequency and the pump frequency, we numerically estimate it by referring to literature experimental data concerning crystalline silicon [39, 40]. We observed that the Raman gain in crystalline silicon presents a Lorentzian profile, centered approximately at 15.6 THz of the pump frequency with a dip located at 16.2 THz [39, 40] and we deduced the Raman peak and dip for the three designed waveguides. For the wCH$_4$ waveguide, with a pump wavelength of 2.100 µm, we can predict that the Raman gain peaks near 2.358 µm and dips around 2.369 µm. In the wNO$_2$ case (2.15 µm pump), the Raman gain peaks around 2.423µm and dips at 2.433µm. Similarly, for the wCOM waveguide, with a pump wavelength of 2.210 µm, the Raman gain peaks near 2.498 µm and dips near 2.510 µm. We can therefore predict a reduced Raman scattering influence, for all three targeted applications, by remembering that for all three waveguides, the targeted signal wavelengths are in the range 3µm-4µm and are therefore well above Raman gain dips.

Considering the mitigation of the Two-Photon Absorption (TPA), the TPA coefficient of crystalline Silicon sharply declines after 1.9µm, while the nonlinear refractive index $n_2$ encounters a peak of high value in between the wavelengths 1.9–2.2 µm [41]. It is worth noting that all our proposed designs exhibit a pump wavelength which falls in the interval of 1.9–2.2 µm. Therefore, TPA is expected to be significantly lower at these pump wavelengths in our designs.

**4.4 Application to gas sensing**

The designed waveguides wCH$_4$ and wNO$_2$ can be incorporated in second order interference experiments [7-9] for obtaining precise gas concentration of CH$_4$ and NO$_2$. The detection of other gases can be achieved by adjusting the core width and/or height of the Silicon waveguide, provided the target gas exhibits an absorption band within silicon's low-loss transparency window ($1.4 - 5.0\mu m$). This capability paves the way for the development of Silicon waveguide arrays, enabling species identification and quantification in gas mixtures, leveraging the simplicity and cost-effectiveness of Silicon waveguide fabrication. In complement to Silicon waveguides, Chalcogenide or Silicon Nitride waveguides can be used to treat wavelengths larger than 5 μm.

**4.5 Application to atmospheric Quantum Key Distribution**

The waveguide wCOM, which is designed to generate the correlated photons with a wavelength of 3.905 μm for the signal photon and an idler photon in the C-band, is relevant to atmospheric daylight QKD. This waveguide may enable a further increase of the wavelength of the photon transmitted in the atmosphere as compared to current Telecom Band systems [2]. Theoretical predictions indicate propagation of MIR signal photons in the atmosphere with lower losses, low solar background noise as compared to Telecom Band systems [4-6].

**4.6 Assets and interest of an integrated platform**

Our silicon waveguides are integrated, but the general set-up is using a $2\mu m$ laser source that is, today, challenging to obtain in a integrated platform. Nonetheless, reader can understand that it is interesting to have an integrated photon pair generation platform, even in the case in which the $2\mu m$ laser source is not integrated, in such a way to reduce the weight and the size overall set-up to some extent, which can be very useful both for gas sensing and for atmospheric QKD. It is interesting to note that current integrated laser sources are offering output peak power of approximately 1mW to 3.5mW [42] and are in constant development, which make us optimistic about their capability to reach the level of peak pump power required by our proposed designed (in between 10mW to 35mW).

## 5. Conclusion

We present three all-solid Silicon waveguides designs for the SFWM generation of correlated photons pairs bridging 3μm-4μm MIR Band and Telecom Band. For deriving these designs, we used a literature model, that we experimentally validated employing results previously reported in the literature. We found very high agreement between theoretical and experimental results.

    The three designs are intended to be applied to gas sensing or MIR daylight atmospheric QKD. Targeting a probability of photon pair generation per pulse of 0.05, we found out the requirement of a pump peak power inferior to 32.2 mW in the case of 5ps secant hyperbolic pulses. The maximum simulated wavelength gap between the idler and signal wavelength in our three waveguides reaches 2364nm, which is dramatically larger than the values theoretically reported with SFWM in all-solid SOI in the case of a quantum regime in which at most one photon pair is emitted per pulse. The corresponding design allows to reach a MIR wavelength as large as 3.905μm.

    By analysis found on previous experimental observations, we predict reduced Raman scattering and Two Photon Absorption and tolerance to small fabrication imperfections. The significant practical advantages of Silicon waveguides make the route proposed here highly attractive for overcoming the current limitations of Spontaneous Parametric Down Conversion realized in high $\chi^2$ platforms, which are costly, present a reduced versatility and are challenging to obtain in integrated platforms.

**<span style="color:red">Data availability statement</span>**

All data that support the findings of this study are included within the article.

## Acknowledgments
The authors would like to thank Indian Institute of Technology Delhi for providing the access to COMSOL Multiphysics 6.2 software suite. AKP and DJ acknowledge the fund support from ANRF's SRG grant.

## Conflict of interest
The authors declare no conflicts of interest.